\documentclass[aps,pre,twocolumn,showpacs]{revtex4}
\usepackage{epsfig}
\usepackage{times}
\usepackage{color, soul}
\begin{document}

\title{Diverging fluctuations in a spatial five-species cyclic dominance game}
\author{Jeromos Vukov,$^1$ Attila Szolnoki,$^1$ and Gy{\"o}rgy Szab{\'o}$^{1,2}$}
\affiliation{$^1$ Institute of Technical Physics and Materials Science, Research Centre for Natural Sciences, Hungarian Academy of Sciences, P.O. Box 49, H-1525 Budapest, Hungary\\
$^2$ Roland E{\"o}tv{\"o}s University, Regional Knowledge Centre, Ir{\'a}nyi D{\'a}niel u. 4, H-8000 Sz{\'e}kesfeh{\'e}rv{\'a}r, Hungary }

\date{\today}
\begin{abstract}
A five-species predator-prey model is studied on a square lattice where each species has two prey and two predators on the analogy to the Rock-Paper-Scissors-Lizard-Spock game. The evolution of the spatial distribution of species is governed by site exchange and invasion between the neighboring predator-prey pairs, where the cyclic symmetry can be characterized by two different invasion rates. The mean-field analysis has indicated periodic oscillations in the species densities with a frequency becoming zero for a specific ratio of invasion rates. When varying the ratio of invasion rates, the appearance of this zero-eigenvalue mode is accompanied by neutrality between the species associations. Monte Carlo simulations of the spatial system reveal diverging fluctuations at a specific invasion rate, which can be related to the vanishing dominance between all pairs of species associations.
\end{abstract}


\maketitle

\section{Introduction}
\label{intro}

The analysis of competing species has a long history in the study of ecological systems \cite{lotka_pnas20, volterra_31, may_74, frey_pa10}. Even if the number of participating species is low, the complex interactions in an ecosystem usually create an ever changing spatial pattern driven by the chaotic nature of the system. Cyclically dominant interaction systems are of great importance in nature \cite{sinervo_n96, kerr_n02, kirkup_n04, kim_pnas08, guill_jtb11} hence they have attracted significant interest from the scientific community as well \cite{masuda_pre06, laird_an06, reichenbach_prl07, claussen_prl08, laird_e08, jiang_ll_njp09,  peltomaki_pre08, wang_wx_pre10b, andrae_prl10, lamouroux_pre12, avelino_pre12, jiang_ll_pre11, juul_pre13, lutz_jtb13}. The simplest cyclic predator-prey system contains three species and is referred to as the Rock-Paper-Scissors game. Its behavior has been studied intensively and well understood even if it is placed into a spatial environment where species are located on some kind of spatial structure and can only interact with their local neighbors \cite{frean_prsb01,szabo_pre99,reichenbach_n07}. Increasing the number of species in the food web, however, makes the emerging dynamics exponentially more complex. As it was recently shown, the topology of the food web alone cannot determine the outcome of evolution because the strength of the invasion processes plays a decisive role~\cite{szabo_pre08, knebel_prl13}.

In this paper, we analyze the behavior of the spatial so-called Rock-Paper-Scissors-Lizard-Spock (RPSLS) model \cite{kass}. This ecosystem is the generalization of the spatial Rock-Paper-Scissors game \cite{tainaka_jpsj88, tainaka_prl89}, containing five species, each with two prey and two predators forming a double cyclic dominance scenario \cite{laird_jtb09}. To fully describe and understand the general system, one would need to study all possible invasion probabilities for the ten interaction pairs, however this is beyond the actual computational capabilities. To overcome this difficulty, we consider a food web that satisfies a higher level of symmetry as only two invasion rates are distinguished. By using this simplified version, we revisit the model outlined in \cite{kang_pa13} where a site exchange mechanism is also introduced to enhance the validity of mean-field approximations.

After introducing the spatial model, it is investigated in the well-mixed limit using the framework of mean-field approximation. Accordingly, we discuss all the possible stationary and oscillating states and analyze the competitive features among the stationary solutions. These results help to understand the spatial version of the model where the individuals of the species are located on a square lattice and can only interact with their immediate neighbors. In structured population, species can form defensive alliances \cite{szabo_pre01b, szabo_jpa05}, which can be considered as additional higher level ``species'' occupying more space that protect themselves from the invasion of external species or associations. The mean-field results can serve as guidelines to show the direction of invasion between these alliances. The analysis of the invasion velocity between alliances having particular spatio-temporal structures is performed by using Monte Carlo (MC) simulations. Both types of approaches justify that the invasion velocities vanish at a specific ratio of invasion rates where the density fluctuations diverge for sufficiently high value of mixing. In the absence of mixing, the divergence of fluctuation is suppressed and the MC data indicate only a sharp peak in the magnitude of fluctuations when varying the invasion rates.
As we show, the absence or presence of the fluctuations' divergence is related to the invasion velocities between species alliances.

\section{The model}
\label{model}

\begin{figure}[ht]
\centerline{\epsfig{file=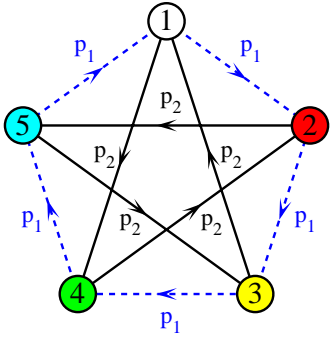,width=4.6cm}}
\caption{(Color online) Food web of cyclic predator-prey system with five species. The directed dashed (blue) and solid (black) lines represent possible invasions.} \label{foodweb5}
\end{figure}

Consider a model with five species cyclically dominating each other according to two invasion routes as illustrated in Fig.~\ref{foodweb5}. To facilitate the notation, we introduce a periodic index for the species where the index $i+1$ subsequent to index $i$ is defined in the following cyclic way: $4 \to 5 \to 1 \to 2 \to 3 \to 4$. The predator-prey interactions are characterized by the following invasion rates: species $i$ replaces species $i+1$ with probability $p_1$ and replaces species $i-2$ with probability $p_2$. These invasion processes are visualized in Fig.~\ref{foodweb5} in the form of a food web where the arrows point from predator to prey.
It is important to note that a species cannot emerge again after extinction due to the defined dynamics.

In the spatial scenario, the individuals of the species are placed on a square lattice of linear size $L$ with periodic boundary condition. The interaction neighborhood of an individual is defined by the four nearest neighbors on the lattice. The size of the lattice is varied between $L=400$ and 2000 depending on the required accuracy of the given MC simulation. In addition to the invasion process, neighboring species may exchange their sites with a probability $p_m$ characterizing the strength of mixing. In an elementary MC step a pair of neighboring individuals is selected randomly from the population that can exchange their position with probability $p_m$, otherwise an invasion takes place between the two individuals with probability $p_1$ or $p_2$ as described above. The simulations are started from a random initial state and the stationary distributions are calculated as long time averages of $5 \cdot 10^6$ MC steps after an initial transitional period of $2 \cdot 10^5$ MC steps. All presented results are averaged over $10 - 30$ independent runs. For the determination of the average invasion velocity between two alliances, however, we used special prepared initial states as detailed later in the paper.

\section{Mean-field analysis}
\label{mfa}

To have a deeper understanding about the behavior of the system, we studied what happens without the spatial constraints in effect. In the well-mixed limit the system can be described by the mean-field approximation. Accordingly, the time dependence of density $x_j$ of species $j$ can be given by the following differential equation system independently of the value of $p_m$:
\begin{equation}\label{definition}
\dot{x}_j = x_j(p_1 x_{j+1} - p_2 x_{j+2} + p_2 x_{j+3} - p_1 x_{j+4}),
\end{equation}
where $j$ runs from 1 to 5 in a cyclic manner.

The sum of all species' densities is 1 by definition and due to the defined dynamics this quantity is conserved. In the general case, when $p_1$ and $p_2$ are positive, the equations can be divided by $p_2$, thus reducing the number of tunable parameters on the right hand side to one: $p_1 / p_2 \equiv q$  , while the left hand side can be kept formally unchanged if we incorporate the factor $p_2$ into the time variable by a rescaling transformation $t'= p_2 t $. The differential equation system can be most effectively analyzed by combining the densities into a vector $\mathbf{x}=(x_1, x_2, x_3, x_4, x_5)$. Using this form, the set of equations can be written as $\dot{x}_j=x_j \mathbf{(Mx)}_j$, where $\mathbf{M}$ is the mean-field interaction matrix given by:
\begin{equation} \mathbf{M} = \left( \begin{array}{rrrrr}
0 & q & -1 & 1 & -q \\
-q & 0 & q & -1 & 1 \\
1 & -q & 0 & q & -1 \\
-1 & 1 & -q & 0 & q \\
q & -1 & 1 & -q & 0 \end{array} \right).
\end{equation}

The antisymmetric structure of the matrix automatically ensures the $\sum\nolimits_{j=1}^5 x_j (t) = 1$ relation. Another constant of motion is $\prod\nolimits_{j=1}^5 x_j$, which is a consequence of the fact that the sum of the elements in each column of $\mathbf{M}$ is zero. The analysis of the matrix' eigenvalue problem [${\bf M}{\bf v}(k)=\lambda(k) {\bf v}(k)$] helps understand both the dynamic and the stationary behaviors of the system. Solving the eigenvalue problem results in the following eigenvectors with complex elements:
\begin{eqnarray}
\mathbf{v}(0) &=& (1,1,1,1,1) \nonumber \\
\mathbf{v}(1) &=& \left( e^{i \varphi},e^{i 2 \varphi},e^{i 3 \varphi},e^{i 4 \varphi},1 \right) \nonumber \\
\mathbf{v}(-1) &=& \left( e^{i 4 \varphi},e^{i 3 \varphi},e^{i 2 \varphi},e^{i \varphi},1 \right)=\mathbf{v^*}(1) \\
\mathbf{v}(2) &=& \left( e^{i 2 \varphi},e^{i 4 \varphi},e^{i \varphi},e^{i 3 \varphi},1 \right) \nonumber \\
\mathbf{v}(-2) &=& \left( e^{i 3 \varphi},e^{i \varphi},e^{i 4 \varphi},e^{i 2 \varphi},1 \right)=\mathbf{v^*}(2) \nonumber \,\,,
\end{eqnarray}
where $\varphi=2 \pi /5$, $i$ is the imaginary unit and asterisk stands for the complex conjugate of the given vector. The numbering of the vectors refers to the wave number of the Fourier components and makes the notation more compact. It is worth noting that the eigenvectors are independent of $q$. In the compact form, element $j$ of the $k$th eigenvector can be given as $v_j(k) =  e ^{i {\varphi}kj}$ where $k = 0, \pm 1, \pm 2$. Accordingly, $v_j(-k)=v_j^*(k)$. Using this notation, the eigenvalues are given as
\begin{equation}
\lambda(k) =  i \omega (k) = 2i [\sin(2 \varphi k)-q \sin (\varphi k)]  \,.
\label{eq:eigval}
\end{equation}
One of the eigenvalues is always zero ($\lambda (0) =0$), while the others are purely imaginary and satisfy the condition $\lambda (-k) =-\lambda (k)$ for $k \ne 0$. In agrement with the expectation, the zero-eigenvalue eigenvector ${\bf v}(0)=(1,1,1,1,1)$ defines the symmetric stationary solution of Eq.~(\ref{definition}) as ${\bf x}(t)={\bf v}^{(12345)}={1 \over 5}{\bf v}(0)$.

\begin{figure}[ht]
\centerline{\epsfig{file=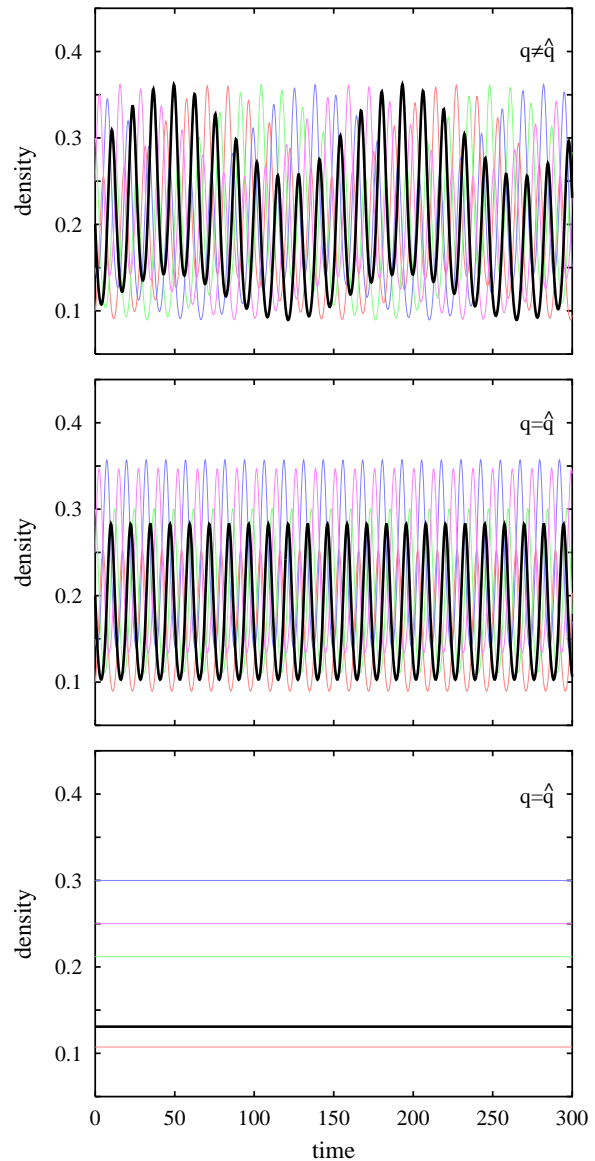,width=7.8cm}}
\caption{\label{fig:meanfield}Time evolution of the species' densities for different invasion rates and initial conditions. \emph{Top panel}: General time evolution; the invasion rates are not at the golden section; the curves are determined by the product of two oscillating periodic functions. \emph{Middle panel}: the invasion rates are fixed to the golden section point but for general initial conditions the time evolution is described by a periodic function determined by the non-zero eigenvalues of the mean-field matrix. \emph{Bottom panel}: the concentrations stay constant when the invasion rates are at the golden section and the initial condition is the linear combination of the zero eigenvalue eigenvectors of the mean-field matrix. For clarity, we have marked one of the curves by thick (black) line.
}
\end{figure}

The other four eigenvectors and eigenvalues determine the system's dynamical behavior if the deviation from the above solution is small, i.e. when the system is started from the close proximity of the symmetric stationary state. In the latter case the solution can be written as the composition of harmonic oscillations around the stationary state:
\begin{equation}
{\bf x}(t)={\bf v}^{(12345)}+  \sum_{k \ne 0} e^{i \omega(k)t} a(k) {\bf v}(k) \,\,.
\label{eq:oscil}
\end{equation}
The solution becomes real if $a(-k)=a^*(k)$. The reader can easily check that in linear approximation ($|a(k)|<<1$) the equation of motion becomes an eigenvalue problem solved by the above expressions. When increasing the deviation from the stationary solution, the harmonic oscillations (composed of two angular frequencies, $|\omega(1)|$ and $|\omega(2)|$) are distorted and shifted in a way satisfying both laws of conservation as illustrated in the top panel of Fig.~\ref{fig:meanfield}.

Besides the above described five-species solutions, the present system has five trivial solutions when only one species remains alive:
\begin{equation}
{\bf x}(t)={\bf v}^{(n)}
\label{eq:homsol}
\end{equation}
with $v_j^{(n)}=\delta_{jn}$ ($j,n=1, \dots , 5$) and $\delta$ denotes the Kronecker's delta function. Two species cannot coexist in the system because the predator prevails over its prey and the whole population terminates in a homogeneous one-species solution. On the other hand, three species can stably coexist if their indices are consecutive and in this case, the dynamics is equivalent to a Rock-Paper-Scissors system. (If there is a gap between the indices, then one of the species has no predator but only two prey and the system evolves into a one-species final state.) Staying at the non-trivial case, the stationary solution for {\it e.g.} species 1, 2, and 3 is given by
\begin{equation}
{\bf x}(t)={\bf v}^{(123)}=\left({q \over 2q+1}, {1 \over 2q +1},{q \over 2q+1},0,0\right).
\label{eq:rsp}
\end{equation}
Moreover, there exist time-dependent solutions for the three species exhibiting periodic oscillations around the stationary solution while two laws of conservation are satisfied, namely, $x_1(t)+x_2(t)+x_3(t)=1$ and $x_1(t)x_2^{q}(t)x_3(t)=$ constant, as detailed in \cite{hofbauer_88, szabo_pr07}. Evidently, four other equivalent three-species solutions can be derived from Eq.~(\ref{eq:rsp}) by shifting the indices cyclically.

The eigenvalues depend on $q$ as defined by Eq.~(\ref{eq:eigval}). The system's behavior changes gradually when two of these eigenvalues vanish. This happens when $\omega(1)=\omega(-1)=0$ at $q=\hat{q}=2 \cos{2 \pi \over 5}={\sqrt{5}-1 \over 2}=0.618034$ (golden ratio). The other possibility [$\omega(2)=\omega(-2)=0$ at $q=-1/\hat{q}$] corresponds to the case when the invasion cycle in the interior of the food web is of reversed direction. This topology can, however, be transformed into the topology of Fig.~\ref{foodweb5} by a suitable rearranging of nodes hence this case is equivalent to the first one.

For $q=\hat{q}$, the finite number of the stationary solutions of Eq.~(\ref{definition}) is extended by a continuous two-dimensional set of solutions derived from the additional zero-eigenvalue eigenvectors ${\bf v}(-1)$ and ${\bf v}(1)$. To simplify the analysis, it is possible to choose two purely real eigenvectors from the two-dimensional subset: $\mathbf{a_1^0} = [{\bf v}(1)+{\bf v}(-1)]/2$, $\mathbf{a_2^0} = [{\bf v}(1)-{\bf v}(-1)]/(2i)$. Any $\mathbf{x}(t)=\frac{1}{5} \mathbf{v}(0)+\alpha \mathbf{a_1^0}+\beta \mathbf{a_2^0}$ linear combination ($0\le x_j \le 1$) of the above vectors results in a stationary distribution (bottom panel of  Figure~\ref{fig:meanfield}). Note that the (eigen)vectors reported in \cite{kang_pa13} are probably mistyped and consequently not correct. The mentioned stationary states form a continuous subset and are asymptotically stable. It is interesting to note that the set contains stationary solutions where only three or four species are present in the population. If the initial population composition is not a pure linear combination of these vectors, then the concentrations oscillate periodically according to the non-zero eigenvalues (see the middle panel of Fig.~\ref{fig:meanfield}). The amplitudes of the oscillations for the different species depend on the actual initial densities. Moreover, there exists a particular set of solutions when the amplitudes are equal for all species.

To complete the mean-field analysis, we examine what happens when the population is started from a state where only one of the species is absent ($x_k=0$ for one of the $1 \le k \le 5$ species). In this case, the course of evolution depends sensitively on the value of $q$. For $q<\hat{q}$ ($q>\hat{q}$), $x_{k+1}$ ($x_{k-1}$) goes to zero in an oscillating way. The closer $q$ to $\hat{q}$ is, the slower the decrease of the given concentration is. If $q$ is exactly at the golden section point then according to the results of the previous paragraph the concentrations of species oscillate periodically or stay constant depending on the composition of the system.

\section{Invasion rates between associations of species}
\label{stability}

In the previous section we have outlined all possible solutions of the equation of motion. In most of the cases Eq.~(\ref{definition}) has 11 stationary solutions. The mean-field analysis has indicated that deviation from the three- and five-species stationary solutions induces oscillations when we modify the densities of species in the initial state. We have to emphasize that the actual spatial systems do not exhibit all the features of the mean-field results. The global oscillations in the species densities can not be observed here because the short range interactions are not capable to synchronize the oscillations that are present locally throughout the whole system. Instead, a self-organizing pattern with moving interfaces and rotating spiral arms can be observed on the two-dimensional lattices. On the contrary, the analysis of the Rock-Paper-Scissors game indicates that global oscillations can be maintained when long-range interactions are introduced in the spatial systems \cite{szabo_jpa04,szolnoki_pre04b}.

Now we discuss what happens when the stationary solutions are considered as associations with the respective composition (or in the case of the spatial model, spatio-temporal structure on the lattice) that compete against each other. For example, the five one-species solutions [given by Eq.~(\ref{eq:homsol})] are not stable because they can be invaded by the offspring of a single individual of their predators. Within the framework of mean-field analysis, the competition between the different associations can be quantified by an average invasion rate measuring the direction of the invasion processes. More precisely, we assume two populations with compositions denoted by ${\bf x}^{\prime}$ and ${\bf x}^{\prime \prime}$ representing two stationary solutions of the equation of motion. The members of these species associations compete in pairs chosen at random from both associations. The mean-field approximation can predict the average invasion rate of association ${\bf x}^{\prime}$ at the expense of ${\bf x}^{\prime \prime}$ with the following vector product:
\begin{equation}
I({\bf x}^{\prime} \to {\bf x}^{\prime \prime})= {\bf x}^{\prime} \cdot {\bf M} {\bf x}^{\prime \prime}.
\label{eq:invrate}
\end{equation}
Evidently $I({\bf x}^{\prime} \to {\bf x}^{\prime \prime})=-I({\bf x}^{\prime \prime} \to {\bf x}^{\prime})$ as ${\bf M}$ is antisymmetric (${\bf M} = - {\bf M}^T$). If the competing associations are one-species solutions given by Eq.~(\ref{eq:homsol}) then $I({\bf v}^{(j)} \to {\bf v}^{(n)})= M_{jn}$. One can easily check that $I({\bf x}^{\prime} \to {\bf x}^{\prime \prime})=0$ if ${\bf x}^{\prime \prime}$ is a one-species solution with the species included in the stationary solution ${\bf x}^{\prime}$. Consequently, $I({\bf v}^{(12345)} \to {\bf v}^{(n)})= 0$ for arbitrary $n$. Similarly, $I({\bf v}^{(123)} \to {\bf v}^{(n)})= 0$ if $n=1, 2$ or $3$ and additional relations can be derived by cyclic permutations. Non-vanishing invasion rates occur when a three-species association competes with another one having one or two common species or with distinct one-species solutions. For example, straightforward calculations give that
\begin{eqnarray}
I({\bf v}^{(123)} \to {\bf v}^{(234)})&=& {-q(1-q-q^2) \over (2q + 1)^2}  \\
I({\bf v}^{(123)} \to {\bf v}^{(345)})&=& {(q-1)(1-q-q^2) \over (2q+ 1)^2} \\
I({\bf v}^{(123)} \to {\bf v}^{(4)})&=& {-(1-q-q^2) \over (2q + 1)}  \\
I({\bf v}^{(123)} \to {\bf v}^{(5)})&=& {(1-q-q^2) \over (2q+ 1)}
\label{eq:inv3s}
\end{eqnarray}
and similar expressions can be derived by cyclic permutation of the indices. Notice that all quantities become zero at the golden ratio ($q=\hat{q}$).

Using more sophisticated equations, one can even evaluate the actual evolution of the composition for both competing associations. Instead of performing such a calculation, we emphasize that $I({\bf x}^{\prime} \to {\bf x}^{\prime \prime})$ characterizes the initial speed of invasion and the meaning of this quantity can be interpreted for the spatial systems, too. For example, when the whole battlefield of the two-dimensional spatial system is divided into two regions and each region is occupied by a mix of species according to the composition of the stationary states ${\bf x}^{\prime}$ and ${\bf x}^{\prime \prime}$ then $I({\bf x}^{\prime} \to {\bf x}^{\prime \prime})$ is proportional to the initial invasion speed as the dynamical process is controlled by short range interactions.

\begin{figure}[ht]
\centerline{\epsfig{file=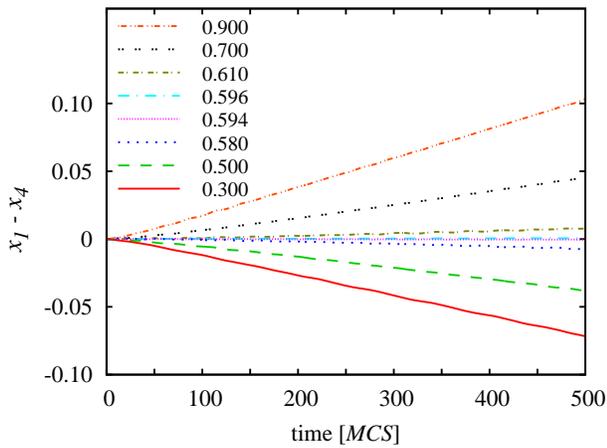,width=8.0cm}}
\caption{(Color online) Difference in species densities $x_1(t)-x_4(t)$ for different $q$ values as a function of time when the three-species solutions ${\bf v}^{(123)}$ and ${\bf v}^{(234)}$ compete along vertical interfaces on a square lattice at $p_m=0.7$. Every curve is an average of 100 independent runs at $L=2000$ system size.}
\label{fig:inv2d}
\end{figure}

\begin{figure}[ht]
\centerline{\epsfig{file=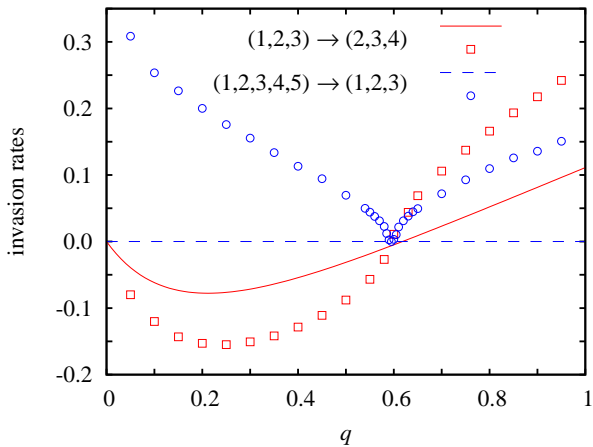,width=8.0cm}} \caption{(Color online) MC results  for the average invasion  velocities between the phases ${\bf v}^{(123)}$ versus ${\bf v}^{(234)}$ (squares) and ${\bf v}^{(12345)}$ versus ${\bf v}^{(123)}$ (circles) as a function of $q$ at $p_m=0.7$. Lines illustrate the mean-field predictions between two stationary solutions as indicated in the legend.} \label{fig:compinv}
\end{figure}

At the same time, for the lattice systems, the invasion process can be investigated by MC simulations as well. In this case, the velocity of invasion can be evaluated conveniently by recording the time-dependence of species densities as illustrated in Fig.~\ref{fig:inv2d}. In the MC simulations, the populations in the two regions are thermalized first: they are started from a state where the individuals are placed on both halves of the lattice according to the densities present in $\mathbf{x}^{\prime}$ and ${\bf x}^{\prime \prime}$ and are allowed to evolve till they reach the corresponding stationary states. At this point, the separating interface is removed and the invasion events are monitored. (Similar technique was already applied in \cite{szolnoki_pre11b,roman_pre13}.)
The invasion velocity between the phases ${\bf v}^{(123)}$ and ${\bf v}^{(234)}$ can be estimated from the change in $x_4(t)-x_1(t)$ averaged over several runs on a sufficiently large lattice. The results obtained for different $q$ values are compared with the theoretical mean-field results in Fig.~\ref{fig:compinv} illustrating the $q$-dependence of the invasion rates. The figure shows that the mean-field approximation predicts the invasion velocity function qualitatively well, namely, the general trend of the curve is in good agreement with the MC data, moreover, it predicts the value of the critical point even quantitatively well. According to the simulations the average invasion velocity becomes zero at $q=q_c=0.596(2)$, a value that is slightly smaller than $\hat{q}$. The zero value of the average invasion velocity can be interpreted as a type of neutrality between the given phases. In this case, the motion of the interface becomes random.

In contrary to the mean-field predictions, the MC simulations indicated a completely different behavior when the phases ${\bf v}^{(123)}$ and ${\bf v}^{(345)}$ are confronted in the same way. More precisely, the five species state (${\bf v}^{(12345)}$) is formed along the interface and spreads with an average velocity that can be quantified, too. Finally it takes over the whole spatial system independently of the value of $q$. Evidently, the average invasion velocity can be well described by the one obtained from the competition of phases ${\bf v}^{(12345)}$ and ${\bf v}^{(123)}$. Here, it is worth emphasizing the technical difficulties caused by the diverging fluctuations in the vicinity of $q_{c}$ where the invasion velocities vanish. Due to these difficulties, henceforth our numerical analysis is restricted to two values of mixing ($p_m=0.7$ and 0) representing different behaviors.

The spatial effects cause similar phenomenon when a three-species phase ({\it e.g.} ${\bf v}^{(123)}$) is confronted with a homogeneous one (${\bf v}^{(n)}$). If $n=1$, 2, or 3 then the homogeneous state is invaded by the corresponding predator, that is invaded by its predator along a second invasion front, and the latter one is also invaded by the third member of the three-species association. After a short period, phase ${\bf v}^{(123)}$ will dominate the invaded territory completely. Similar transient phenomena can be observed for $n=4$ (or 5). However, in this case, additional three-species associations are formed (${\bf v}^{(234)}$ or ${\bf v}^{(512)}$) that can spread over the territory of the homogeneous phase while it competes with the original three-species phase. The final stationary state is determined by the competition between the two three-species phases as described above. Similar phenomena were already reported for other types of spatial predator-prey models \cite{szabo_pre08b}.

\begin{figure}[ht]
\centerline{\epsfig{file=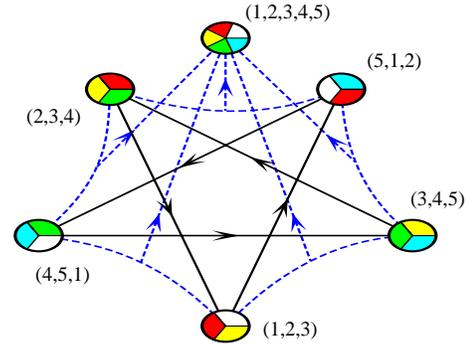,width=6.0cm}}
\caption{(Color online) Flow diagram without the five homogeneous states for $q>q_c$. The invasion processes between the three-species solutions (nodes with three colors) are indicated by directed solid lines. Dashed (blue) lines refer to transitions when the confrontation of two three-species solutions leads to the emergence of the symmetric five-species phase. Note that arrows point towards the winning associations.} \label{foodweba}
\end{figure}

All the above-mentioned phenomena can be summarized in a flow diagram where the nodes represent the possible 11 stationary solutions. In comparison to the food web (see Fig.~\ref{foodweb5}),
the resultant flow diagram involves transitions where the confrontation of two states results in a third one. The mentioned transitions are displayed in Fig.~\ref{foodweba} where the homogeneous states are omitted to avoid confusion.
Notice that the direction of the arrows along the pentacle refers to the cases when $q > q_c(p_m=0.7)=0.596(2)$. In the opposite case ($q < q_c(p_m=0.7)$) the directions of invasion between the three-component associations, marked by solid (black) lines in Fig.~\ref{foodweba}, are reversed as shown in Fig.~\ref{fig:compinv}. We should stress that the direction of invasions between associations can change without changing the direction of invasions between the single species (summarized in Fig.~\ref{foodweb5}). In other words, this system gives another example when the topology of the food web cannot unambiguously determine the outcome of the evolutionary process.

\begin{figure}[ht]
\centerline{\epsfig{file=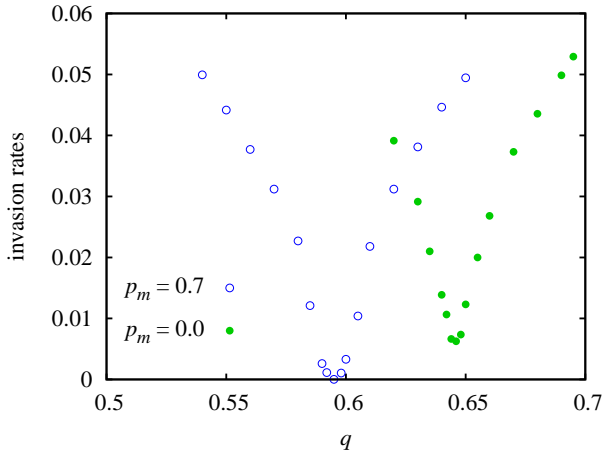,width=8.0cm}} \caption{(Color online) Average invasion velocities between the five- (${\bf v}^{(12345)}$) and three-species states (${\bf v}^{(123)}$) for $p_m=0.7$ (blue open circles) and $p_m=0$ (green closed circles).}
\label{fig:invx}
\end{figure}

The curious feature of the present system is that the average invasion velocity between the five-species solution (${\bf v}^{(12345)}$) and any other three-species solution becomes zero at $q=q_c$ within the statistical error for $p_m=0.7$ as illustrated in Fig.~\ref{fig:invx}. On the contrary, for $p_m=0$ the same invasion velocity remains positive and exhibits a minimum in the vicinity of $q=0.648(4)$.

Now we discuss the consequences of the situations when all the average invasion velocities (between the species associations) vanish simultaneously. In the spatial systems the concept of species associations becomes relevant at a coarse-grained level when the lattice is divided into cells containing a sufficiently large number of lattice points providing the maintenance of the three-species states for a long time ({\it e.g.} $50 \times 50$ or $100 \times 100$ used in previous simulations \cite{kang_pa13}). Within such small isolated cells, the system develops fast into states where some of species are absent. Due to the microscopic interactions (mixing and invasion) the neighboring cells influence each other in a way described above. However, if all the average invasion velocities are zero then the system's behavior can be well approximated by a voter model \cite{liggett_85}. Accordingly, the two-dimensional system is expected to evolve very slowly towards one of the homogeneous states \cite{dornic_prl01} or one of the three-species associations as reported by Kang {\it et al.} \cite{kang_pa13}. This slow ordering process is blocked by the presence of cyclic dominance among the species associations as it was observed for some other systems \cite{szabo_jtb07,szabo_pre08b}. In the latter cases, rotating spiral arms can be observed between the states dominating each other cyclically independently of the fact that they are composed of a single or several species \cite{szabo_pr07}. The rotating spiral arms become striking when the interfacial regularities are reduced by taking the Potts energy into consideration \cite{szolnoki_pre05} or by dividing the invasions into two consecutive elementary processes with the introduction of empty sites \cite{reichenbach_n07, peltomaki_pre08, jiang_ll_pla12, wang_wx_pre10b, yang_r_c10, wang_wx_pre11, rulands_pre13, feng_ss_pa13}.

The simplest model to study the transition between the mentioned behaviors was suggested by Tainaka and Itoh \cite{tainaka_epl91} who modified a three-state voter model with introducing an additional cyclic dominance. This model is equivalent to a three-species cyclic predator-prey (or Rock-Paper-Scissors) model where both directions of invasion are allowed with probabilities $p$ and $(1-p)$. For $p=1/2$, this model becomes equivalent to the voter model, otherwise all three species remain alive and the resultant self-organizing spatiotemporal pattern can be characterized by the density of the rotating spirals (called vortices). The numerical investigations of this model indicated that the density of vortices goes to zero algebraically when $p \to 1/2$ \cite{tainaka_epl91, szabo_pre99}. At the same time both the correlation length (or the average distance between the rotating vortices) and the magnitude of fluctuations diverge when approaching the critical point known exactly here ($p=1/2$). In the present system the investigation of the fluctuations seems to be the most convenient way to check the validity of the above picture.

\section{Fluctuations}
\label{fluctuations}

The fluctuation of species densities is measured by the quantity
\begin{equation}
\chi= {N \over 5} \sum_j \left \langle \left( x_j(t)-{1 \over 5} \right)^2 \right \rangle \,
\label{eq:chi}
\end{equation}
where $\langle \ldots \rangle$ denotes averaging over a sufficiently long period of time. This quantity becomes independent of the players' number $N=L^2$ if the linear size of the system is significantly larger than the longest correlation length in the spatial distribution. Similar quantity is widely used for the classification of the critical phase transitions \cite{stanley_71}.

\begin{figure}[ht]
\centerline{\epsfig{file=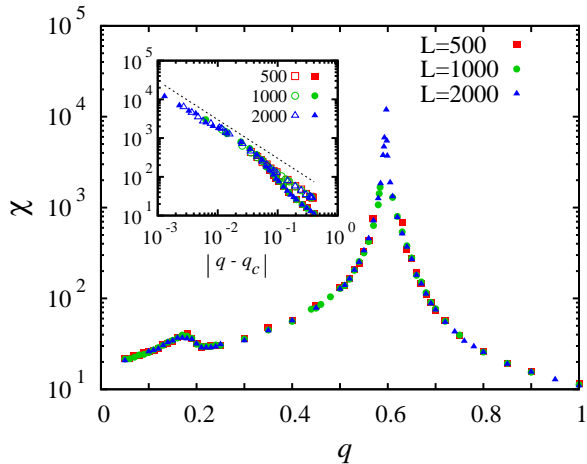,width=8cm}} \caption{(Color online) Fluctuation $\chi$ in dependence of $q$ at $p_m=0.7$ migration rate. Data obtained for different linear size are indicated by colored symbols as explained in the legend. The fluctuations diverge at critical point $q_c=0.5957(3)$. Inset shows power-law behavior with an exponent $\gamma=1.0(1)$, marked by dotted line. In the inset, open (closed) symbols represent $\chi$ values obtained below (above) the critical point.}
\label{fig:fluct}
\end{figure}

We have measured $\chi$ at different values of $q$ at fixed $p_m=0.7$ migration rate. The presented results, summarized in
Fig.~\ref{fig:fluct}, were obtained as an average of 20 independent runs up to $5 \cdot 10^6$ MC steps at every $q$ value for linear sizes of $L=500$, 1000, and 2000. Figure~\ref{fig:fluct} illustrates that there are two peaks at two different values of $q$. While scaled fluctuations remain finite at the smaller $q$, $\chi$ will diverge in an infinite system at a critical point of $q_c=0.5957(3)$. This is confirmed by the inset where the MC data are fitted to the
\begin{equation}
\chi \propto |q - q_c|^{-\gamma}
\label{eq:gamma}
\end{equation}
scaling law. The fitted critical exponent is $\gamma=1.0(1)$. Naturally, in the vicinity of the critical point the amplitude of oscillations can be so large in a finite system that some species become extinct resulting in a three-species final state, as it was
observed in \cite{kang_pa13}.
In fact there exists a natural upper limit for the fluctuation as $\chi_{ul}= {N \over 5} \sum_j \langle(v^{(123)}_j-{1 \over 5})^2\rangle \simeq 0.03 N$ characterizing $\chi$ when the system evolves into one of the three-species stationary solutions. Consequently in Fig.~\ref{fig:fluct} we have plotted only those Monte Carlo data that satisfy the conditions $\chi << \chi_{ul}$ ensuring the survival of all the five species in the system during the simulations.

Additionally, we have tested the divergence of fluctuations in a voter model-like five-species predator-prey model where along the outer edges of the food web (see Fig.~\ref{foodweb5}) both the forward and backward invasions are allowed with probabilities $1/2 \pm r$ while a mixing with $p_m=1/2$ is also allowed as above. The preliminary results support the above mentioned picture but the detailed analysis goes beyond the scope of the present work.

\begin{figure}[ht]
\centerline{\epsfig{file=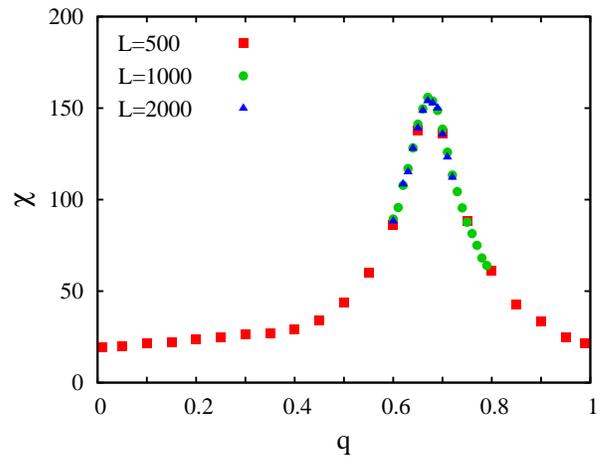,width=8cm}}
\caption{(Color online) Monte Carlo data for $\chi$ versus $q$ in the absence of migration. }
\label{fig:chi0}
\end{figure}

Besides the power-law divergence, the readers can observe a smaller peak in $\chi$ at $q \simeq 0.177(3)$. This mysterious peak also appears in the $\chi$ {\it vs.} $q$ function for other values of mixing and both its position and its height depend on $p_m$. It is worth mentioning that the choice of $p_m=0.7$ was motivated by the demand of quantifying this peak as accurately as possible. For higher values of $p_m$, $\chi$ increases in the vicinity of this peak and this fact causes technical difficulties in its accurate analysis. In the opposite case ($p_m \to 0$) the peak vanishes (see Fig.~\ref{fig:chi0}). It is conjectured that the emergence of this peak can be related to the formation of a spatio-temporal pattern where the cyclic dominance among the three-species association leads to the formation of a more complicated structure that can be characterized by another (longer) correlation length.

In the absence of mixing ($p_m=0$) the MC simulations indicate a significantly different behavior in the fluctuations. Instead of the power law divergence, a sharp but finite peak can be observed in $\chi$ at $q_c(p_m=0)=0.675(3)$ as illustrated in Fig.~\ref{fig:chi0}.
In this case, the two relevant invasion velocities do not vanish simultaneously invalidating the analogy to the voter model at the level of competition between the three-species associations. The low values of invasion velocities, however, support the formation of large domains of three-species associations and it is usually accompanied by an enhancement in $\chi$, too \cite{stanley_71}.

\section{Summary}
\label{sec:summary}

We have reinvestigated the spatial Rock-Paper-Scissors-Lizard-Spock model independently studied by Hawick \cite{CSTN-129} and Kang {\it et al.} \cite{kang_pa13} where the latter group distinguished only two types of invasion rates ($p_1$ and $p_2$) while the neighboring species were allowed to exchange their position with a probability $p_m$ on the square lattice. Varying the ratio of the invasion rates ($q=p_1/p_2$) we have analyzed the behavior of this model by using mean-field approximations and performing MC simulations. Within the framework of mean-field approximation, this system has 11 stationary solutions that can be considered as species associations playing relevant roles in the behavior of the spatial system. The analysis of the time-dependent solutions of the mean-field equations has indicated the presence of two harmonic oscillations in the vicinity of the symmetric five-species stationary solution. The frequencies of the harmonic oscillations are directly related to the eigenvalues of the matrix ${\bf M}$ defining the bilinear equations of motion. The analytical calculations indicate that these eigenvalues depend on $q$ and two of them vanish at the golden ratio ($q=\hat{q}=(\sqrt{5}-1)/2$). The corresponding zero-frequency mode expands the set of the stationary solutions at $q=\hat{q}$ and is responsible for the observed anomalies. Here it is worth mentioning that very recently similar zero-eigenvalue strategies were described by Press and Dyson \cite{press_pnas12}, who studied evolutionary prisoner's dilemma game with stochastic reactive strategies \cite{nowak_amc89, nowak_jtb89}, that can be utilized for different purposes.

In the present system it is found that the competition (dominance) between the mentioned species associations can be characterized by a simple quantity related to ${\bf M}$ and the possible solutions of the mean-field approximations. This analysis indicates vanishing dominance between any pair of mean-field solutions at the golden ratio. At this point, the direction of invasions between associations is reversed showing that the topology of the food web does not always determine the final state of the evolution. Although the spatial effects modify the dominance between the structured species associations, some relations remain valid if the site exchange mechanism (mixing) is sufficiently intensive. Namely, dominance between relevant species associations can vanish simultaneously when tuning the value of $q$ and this event is accompanied by a divergence in the species density fluctuations quantified by $\chi$. According to the MC simulations, the latter effect occurs at $q=q_c \ne \hat{q}$ depending on $p_m$. In the absence of mixing, the dominance between the pairs of species associations does not vanish simultaneously and only a sharp peak can be observed in $\chi$ instead of its power law divergence. The mentioned features can be well approximated by simple models based on the voter model combined with cyclic dominance.

It is emphasized that for sufficiently large fluctuations, two of the five species can become extinct within a short transient time and the system evolves into one of the three-species solutions if the linear size is not large enough. The latter finite size effect, observed in \cite{kang_pa13}, can even occur in the absence of the divergence of $\chi$.

At the microscopic level, the spatio-temporal evolution of the species distribution is controlled by elementary invasions and site exchange between the neighboring lattice sites. For larger length scales, the behavior of these systems can be described via introducing similar interactions between species associations consisting of one or more species. In this case, we can introduce a generalized "food web" displaying the new invasion processes. Evidently, this approach implies the possibility for the emergence of a hierarchy of species associations as it occurs in nature.

\section*{Acknowledgements}

This work was supported by the John Templeton Foundation (FQEB Grant \#RFP-12-22), the Hungarian National Research Fund (OTKA TK-101490), and the European Social Fund through project FutureICT.hu (T\'AMOP-4.2.2.C-11/1/KONV-2012-0013).

\end{document}